\documentclass[preprint,amsmath,amssymb]{revtex4}
\usepackage{dcolumn}
\usepackage{bm}
\usepackage{graphicx}
\usepackage{subfigure}
\usepackage{color}

\newcommand{\be}{\begin{eqnarray}}
\newcommand{\ee}{\end{eqnarray}}

\begin{document}

\title{Topology of the landscape and dominant kinetic path
for the thermodynamic phase transition of the charged Gauss-Bonnet AdS black holes}

\author{Ran Li$^{a}$}
%\thanks{liran@htu.edu.cn}

\author{Conghua Liu$^{b,c}$}
%\thanks{liuch20@mails.jlu.edu.cn}

\author{Kun Zhang$^c$}
%\thanks{zhangkun@ciac.ac.cn}

\author{Jin Wang$^{d}$}
\thanks{Corresponding author, jin.wang.1@stonybrook.edu}

\affiliation{$^a$ Center for Theoretical Interdisciplinary Sciences, Wenzhou Institute, University of Chinese Academy of Sciences, Wenzhou, Zhejiang 325001, China}

\affiliation{$^b$ College of Physics, Jilin University, Changchun 130022, China}

\affiliation{$^c$ State Key Laboratory of Electroanalytical Chemistry, Changchun Institute of Applied Chemistry, Chinese Academy of Sciences, Changchun 130022, China}

\affiliation{$^d$ Department of Chemistry and Department of Physics and Astronomy, State University of New York at Stony Brook, Stony Brook, New York 11794, USA}

\begin{abstract}
We study the generalized free energy of the five dimensional charged Gauss-Bonnet AdS black holes in the grand canonical ensemble by treating the black hole radius and the charge as the order parameters. On the two dimensional free energy landscape, the lowest points in the basins represent the local stable black holes and the saddle point represents the unstable black hole. We show that black hole is the topological defect of gradient field of the landscape. The black hole stability is determined by the topography of the free energy landscape in terms of the basin depths and the barrier height between the basins and is not by the topology of the gradient field. In addition, we study the stochastic dynamics of the black hole phase transition and obtain the dominant kinetic path for the transition on the free energy landscape. Unlike the one dimensional landscape, the dominant kinetic path between the small and the large black hole state does not necessarily pass through the intermediate black hole state. Furthermore, the inhomogeneity in diffusions can lead to the switching from the coupled cooperative process of black hole phase transition to the decoupled sequential process, giving different kinetic mechanisms.      
\end{abstract}

\maketitle

\section{Introduction}

Recently, the concept of the generalized free energy has been proposed to investigate the thermodynamics and the kinetics of the black hole phase transition \cite{Li:2020khm,Li:2020nsy}. In this aspect, the key ingredient is to introduce the order parameter of the phase transition that describes the microscopic degrees of freedom at the coarse-grained level \cite{Wei:2015iwa,Wei:2019uqg}. It is also shown that the generalized free energy can be derived from the Euclidean gravitational action of the non-equilibrium black holes with the Euclidean conical singularities \cite{Li:2022oup}. This derivation established a concrete foundation of the free energy landscape description of the thermodynamics of the black hole phase transition.

The free energy landscape can be explored to quantify the underlying topography such as the basin depths and the barrier height between the basins of the attraction \cite{GPT,FSW,JW,JWRMP}. From the free energy landscape topography, one can easily read off the stabilities of the on-shell black holes. The reason is that, if there are several on-shell states at a fixed ensemble temperature, the thermodynalically favored state is given by the state with the smallest free energy. Another advantage of free energy landscape quantification is that it allows us to study the kinetics of the phase transition process by treating the landscape as the thermodynamical potential that provides the deterministic driving force and treating the interaction between the black hole system and the bath as the thermal fluctuations that provide the stochastic force. Based on these assumptions, the Markovian dynamics and the non-Markovian dynamics of the black hole phase transition have been investigated \cite{Li:2021vdp,Li:2022ylz,Li:2022yti}.

In the previous works \cite{Li:2020khm,Li:2020nsy}, the order parameter of the black hole is chosen to be the radius of the event horizon and the generalized free energy is defined as the function of a single variable or the order parameter. Then, the free energy landscape is represented as a one dimensional curve \cite{Wei:2020rcd,Li:2020spm,Wei:2021bwy,Cai:2021sag,Lan:2021crt,Li:2021zep,Yang:2021nwd,Mo:2021jff,Kumara:2021hlt,Li:2021tpu,Liu:2021lmr,Xu:2021usl,Du:2021cxs,Dai:2022mko,Luo:2022gss,Xu:2022jyp}. In general, for other complex systems, for example, proteins in biophysics, chemical reaction systems, et.al. the free energy is the function of multiple order parameters and the landscape is intrinsically high dimensional \cite{JW,JWRMP}. One question naturally raised is whether there is high dimensional landscape to describe the black hole phase transition.

In the present work, we consider the thermodynamic phase transition of the five dimensional charged Gauss-Bonnet black holes \cite{Boulware:1985wk,Cai:2001dz,Wiltshire:1985us,Cvetic:2001bk}. It was shown in \cite{Cai:2013qga,Zou:2014mha} that the small/large black hole phase transition, which exhibits analogy with the Van de Waals liquid-gas system \cite{Kubiznak:2012wp,Kubiznak:2014zwa,Kubiznak:2016qmn}, holds in five-dimensional spherically symmetric charged Gauss-Bonnet-AdS black holes when its potential $\Phi$ is fixed within the range of $0<\Phi<\frac{\sqrt{3}}{4}\pi$. Based on this observation, we work in the grand canonical ensemble and define the generalized free energy as the function of the black hole radius and the black hole charge. The corresponding free energy landscape can then be represented as an intrinsic two dimensional surface. As far as we know, this example is the only case that the two order parameters can be introduced and the free energy landscape is two dimensional.

The black hole states are represented as the points on the landscape, and the on-shell black hole states are the extremum points of the landscape. From the topography of the landscape, one can easily read off the stabilities of the on-shell black holes. We find that the thermodynamically stable Gauss-Bonnet black holes correspond to the lowest points in the sink on the landscape, while the unstable Gauss-Bonnet black hole corresponds to the saddle point. It should be mentioned that, in Ref.\cite{Wei:2022dzw}, by introducing a parameter $\Theta$ and an ancillary vector field $\phi$, the authors argued that the positive and negative winding numbers of the defects correspond to the local thermodynamical stable and unstable black holes. In the present work, we propose that, in order to reveal the relation between the topology of the landscape and the stabilities of the on-shell black holes, one should investigate the intrinsic topological properties of the gradient field of the free energy landscape, even for the one dimensional landscape. We find that the black hole can be treated as the topological defects of the gradient field of the landscape, but there is no indication that the topology is relevant to the stabilities of the on-shell black holes. We emphasize that the stabilities of the black holes is completely determined by the topography of the landscape.

We also study the phenomenological description of the black hole phase transition based on the stochastic dynamics. Considering the black hole as the thermal entity, there should thermal fluctuation and particle number fluctuation in the grand canonical ensemble. Without those fluctuations, the local stable black holes will remain in the basin of the free energy landscape and no phase transition occurs. In analogy to the Van der Waals fluid-gas phase transition \cite{Malakhov:1994}, if these fluctuations are taken into account, the local stable black hole will undergo a stochastic motion on the landscape, which can be effectively described by the Langevin equation for the trajectories or the Fokker-Planck equation for the time evolution of the associated probability equivalently. For the two dimensional free energy landscape as discussed in the present work, one can determine the dominant kinetic path for the stochastic motion on the landscape by using the path integral formalism of the stochastic dynamics. For the one dimensional landscape, the kinetic path between the small black hole state and the large black hole state must pass through the intermediate black hole state \cite{Liu:2021lmr}. By minimizing the abbreviated action functional \cite{Landau,Wangprl2006,Faccioli2006,WangJCP2010}, we show that the dominant kinetic path passes through the intermediate black hole state when the fluctuation is very small while at the finite fluctuations it does not necessarily pass through the intermediate black hole state. Furthermore, we find that the inhomogeneity in diffusions can lead to the switching from the coupled cooperative process of black hole phase transition to the decoupled sequential process, giving rise to different kinetic mechanisms.

This paper is arranged as follows. In Section \ref{landscape}, we defined the generalized free energy of the charged Gauss-Bonnet black holes and discuss the corresponding landscape in the grand canonical ensemble. In Section \ref{stability}, we discuss the stabilities of the on-shell black holes from the topography of the landscape. In Section \ref{topology}, we address the question of how to study the intrinsic topological properties of the landscape. In Section \ref{kinetics}, we study the dominant kinetic path of the phase transition by using the path integral formalism. The conclusion is presented in the last section.

\section{Free energy landscape of Gauss-Bonnet AdS black holes}
\label{landscape}

In this section, we introduce the charged Gauss-Bonnet black holes in five dimensions and summarize the primary thermodynamics properties. Then, we will give the definition of the generalized free energy in the grand canonical ensemble, which is the function of the black hole radius and the charge. The corresponding landscape is also presented.

For our purpose, we consider the spherically symmetric charged Gauss-Bonnet AdS black hole in $D=5$ dimensions that is described by the line element \cite{Boulware:1985wk,Cai:2001dz,Wiltshire:1985us,Cvetic:2001bk}
\begin{eqnarray}
ds^2=-f(r)dt^2+\frac{dr^2}{f(r)}+r^2 d\Omega_3^2\;, 
\end{eqnarray}
where the metric function $f(r)$ is given by 
\begin{eqnarray}
f(r)=1+\frac{r^2}{2\alpha}\left(1-\sqrt{1+
\frac{32\alpha M}{3\pi r^4}- \frac{4\alpha Q^2}{3r^6} -\frac{16\pi\alpha P}{3}}\right)\;,
\end{eqnarray}
and the line element of the unit 3-sphere $S^3$ in hyperspherical coordinates $(\psi,\phi,\varphi)$ is given by
\begin{eqnarray}
d\Omega_3^2=d\psi^2+\sin^2 \psi\left(d\theta^2+\sin^2\theta d\varphi^2\right)\;.
\end{eqnarray}
For the hyperspherical coordinates $(\psi,\phi,\varphi)$, $\psi$ and $\theta$ run over the range $0$ to $\pi$, and $\phi$ runs over $0$ to $2\pi$. The volume of the hypersphere $S^3$ is $2\pi^2$.

In the expression of the metric function $f(r)$, the parameter $M$ is the black hole mass, the parameter $Q$ is the charge of the black hole, and the pressure $P=-\Lambda/8\pi$ is defined in terms of the cosmological constant \cite{Kastor:2009wy,Dolan:2011xt}. It should be noted that, in order to have a well-defined vacuum in the theory, the effective Gauss-Bonnet coefficient $\alpha$ should satisfy the constraint \cite{Cai:2013qga,Zou:2014mha}
\begin{eqnarray}
0\leq \frac{64\pi \alpha P}{(D-1)(D-2)}\leq 1\;. 
\end{eqnarray}

The thermodynamics of this black hole has been comprehensively discussed \cite{Cai:2013qga,Zou:2014mha} . Let us summarize the primary conclusions. The mass, the Hawking temperature, the entropy, and the electric potential of the event horizon can be expressed in terms of the black hole radius $r_h$ as \cite{Cai:2013qga,Zou:2014mha} 
\begin{eqnarray}\label{expressions}
M&=&\frac{3}{8} \pi  r_h^2 \left(1+\frac{\alpha}{r_h^2}+ \frac{4}{3} \pi  P r_h^2\right)+\frac{\pi  Q^2}{8 r_h^2}\;,
\nonumber\\
T_H&=&\frac{3r_h^4+8\pi P r_h^6-Q^2}{6\pi r_h^3(r_h^2+2\alpha)}\;,
\nonumber\\
S&=&\frac{\pi^2}{2}r_h(r_h^2+6\alpha)\;, 
\nonumber\\
\Phi_H&=&\frac{\pi Q}{4r_h^2}\;.
\end{eqnarray}
In these expressions, we can observe that these thermodynamic quantities can be considered as the function of the black hole radius $r_h$ and the charge $Q$. It has been shown that, when $0<\Phi_H<\frac{\sqrt{3}}{4}\pi$, there exists the small/large black hole phase transition in the grand canonical ensemble \cite{Cai:2013qga,Zou:2014mha}. In the following, we will investigate the phase transition in the framework of the free energy landscape.

We now define the generalized free energy function for the charge Gauss-Bonnet black holes in the grand canonical ensemble as \cite{York:1986it,Whiting:1988qr,Braden:1990hw}
\begin{eqnarray}
F(r_h,Q)&=&M(r_h,Q)-T S(r_h)-Q \Phi\nonumber\\
&=& \frac{3}{8} \pi  r_h^2 \left(1+\frac{\alpha}{r_h^2}+ \frac{4}{3} \pi  P r_h^2\right)+\frac{\pi  Q^2}{8 r_h^2}- \frac{\pi^2}{2}r_h(r_h^2+6\alpha) T-Q\Phi\;. 
\end{eqnarray}
Note that in the definition of the generalized free energy, the two parameter $T$ and $\Phi$ are considered as the ensemble parameter. This is to say that $T$ and $\Phi$ are the temperature and the chemical potential of the thermal and the particle bath that the system contacts with. They are adjustable and determined by the external environment or the bath. Therefore, in the grand canonical ensemble, the free energy function $F(r_h,Q)$ is defined as the function of black hole radius $r_h$ and charge $Q$ and the corresponding landscape is intrinsically two dimensional.

In Fig.1, the two dimensional free energy landscape for the charged Gauss-Bonnet AdS black hole is plotted. In this plot, we have chosen the ensemble temperature $T$ and the chemical potential $\Phi$ to be the same as the critical temperature $T_c$ and the critical chemical potential $\Phi_c$ of the small/large black hole phase transition. We show that the landscape has two separated basins and the basins are connected by a saddle point. At the critical point of phase transition, the two basins have the same depths. Otherwise, the two basins have different depths depending on the ensemble temperature and the chemical potential.

\begin{figure}
  \centering
  \includegraphics[width=8cm]{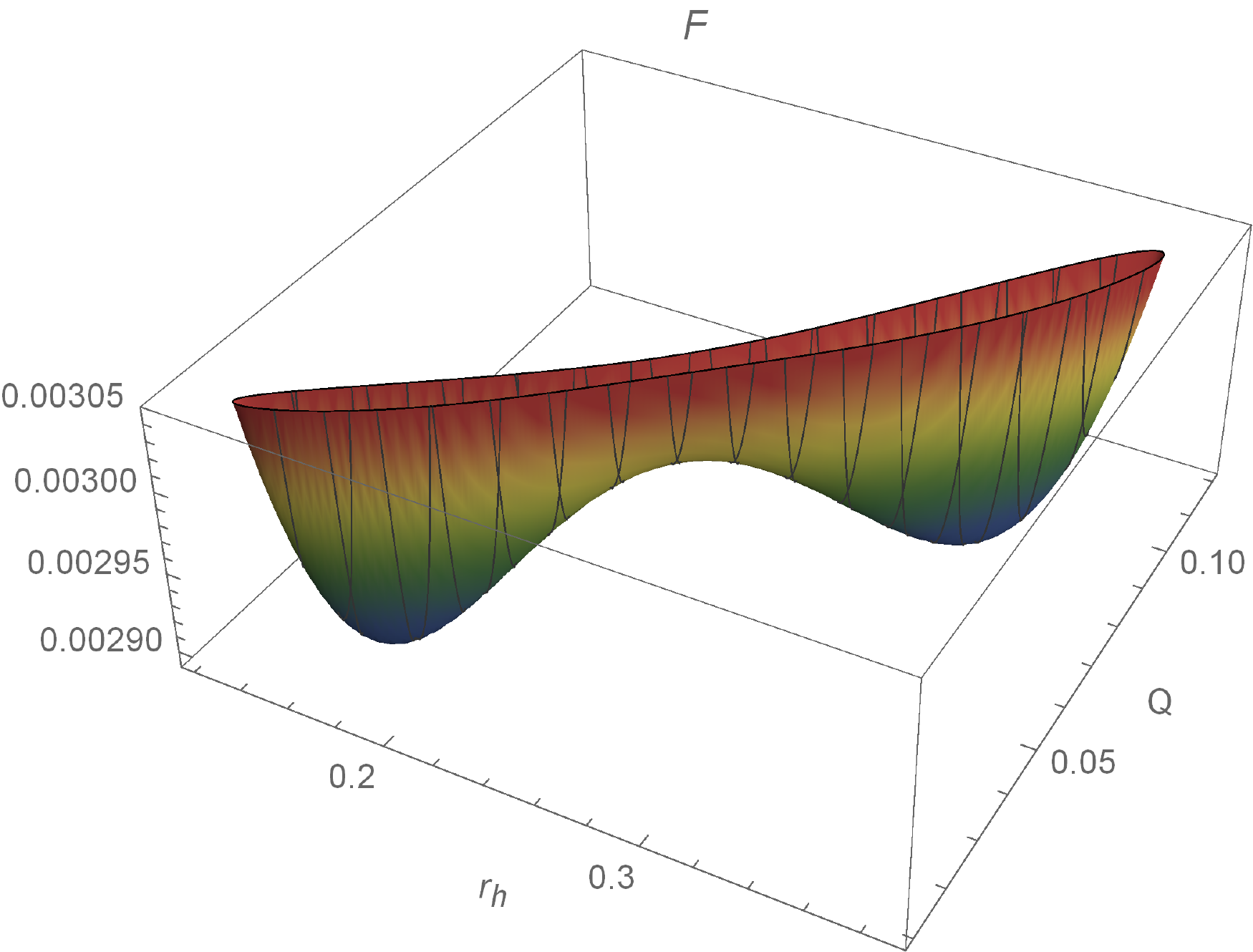}\\
  \caption{Two dimensional free energy landscape for the charged Gauss-Bonnet AdS black holes. The ensemble temperature and chemical potential are taken as the critical temperature and chemical potential of the small/large black hole phase transition. The two basins have the same depths. In this plot, $\alpha=0.01$, $P=0.5$, $T=0.514$, and $\Phi=0.6$. }
  \label{Free_Energy_Landscape}
\end{figure}

\begin{figure}
  \centering
  \includegraphics[width=8cm]{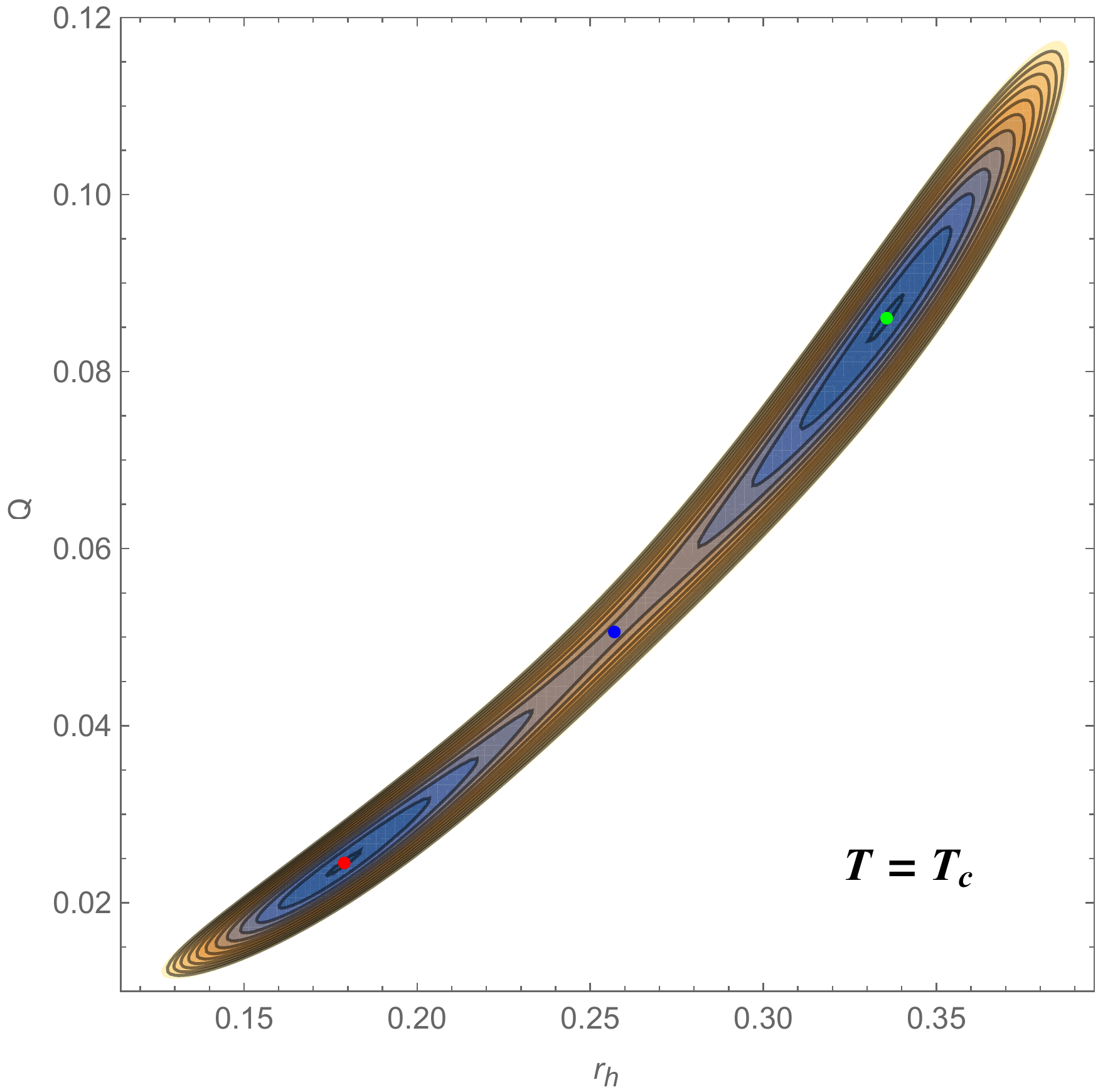}
  \includegraphics[width=2cm]{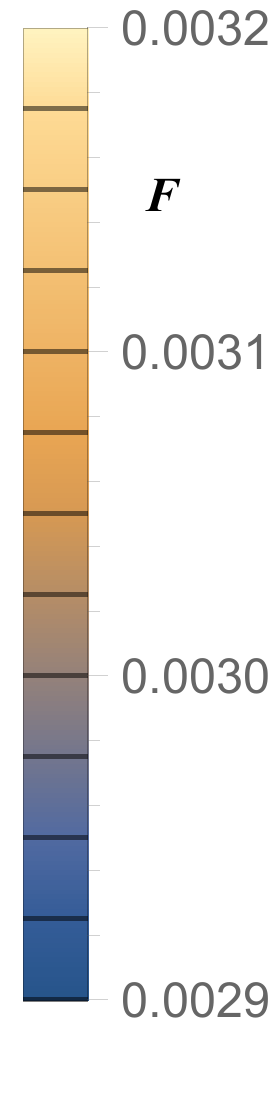}\\
  \caption{The contour plot of the two dimensional free energy landscape in Fig.\ref{Free_Energy_Landscape}. The red, the blue, and the green points represent the small, the intermediate, and the large Gauss-Bonnet AdS black holes. }
  \label{Density_Plot}
\end{figure}

The corresponding two dimensional contour plot of the generalized free energy is presented in Fig.2. In this plot, there are three special points, i.e. the two lowest points in each basin and the saddle point, which are denoted by different colours. They are the extremum points on the landscape, which are determined by the necessary conditions for the extremum of the free energy function
\begin{eqnarray}\label{Equilibirum_con}
\frac{\partial F}{\partial r_h}=\frac{\partial F}{\partial Q}=0\;.
\end{eqnarray}
It is easy to verify that the two conditions give rise to the expressions of the Hawking temperature $T_H$ and the electric potential $\Phi_H$ in Eq.(\ref{expressions}). This means that the three extremum points on the landscape represent the three branches of on-shell black hole solutions, and their Hawking temperature as well as the electrical potential are equal to those of the ensemble temperature and potential. The three branches of black holes are in equilibrium state with the bath. In this way, the necessary conditions (\ref{Equilibirum_con}) for the extremum are also the equilibrium conditions for the on-shell black holes.

\section{Stability of black hole from the topography of free energy landscape} 
\label{stability}

In this section, we discuss the stabilities of the small/intermediate/large Gauss-Bonnet black holes from the topography of free energy landscape.

The generalized free energy provides us not only the free energies of the on-shell black holes, but also the free energies of the off-shell black holes. As mentioned, the equilibrium conditions Eq.(\ref{Equilibirum_con}), i.e. the extremum conditions for the landscape, lead to the expressions of the Hawking temperature and the electric potential of the on-shell black hole solutions. This is to say that the extremum points on the landscape just represent the on-shell black hole solutions, while other points represent the off-shell black hole states. From Fig.\ref{Free_Energy_Landscape} and Fig.\ref{Density_Plot}, we observe that the small, and the large black holes are represented by the lowest points of the basins on the landscape, and the intermediate black hole is represented by the saddle point.

In the plots of Fig.\ref{Free_Energy_Landscape} and Fig.\ref{Density_Plot}, the two basins have the same depths because the external parameters are selected to be the critical parameters of the black hole phase transition. Then, in this case, the small and the large black holes have the same on-shell free energies while the intermediate black hole has the higher free energy. Obviously, the small, and the large Gauss-Bonnet black holes are thermodynamically stable and the intermediate black hole is unstable.

\begin{figure}
  \centering
  \includegraphics[width=8cm]{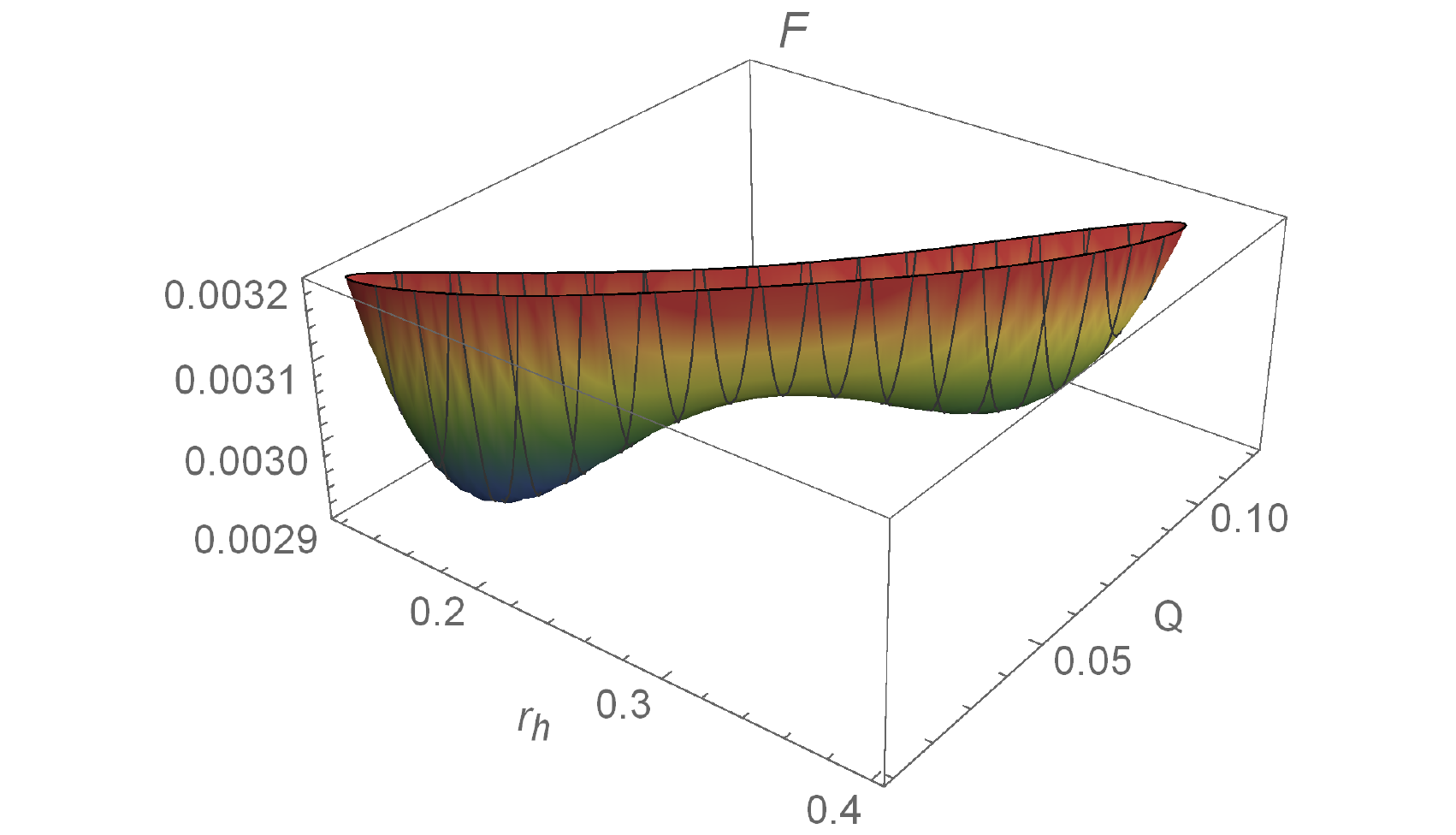}\\
  \includegraphics[width=8cm]{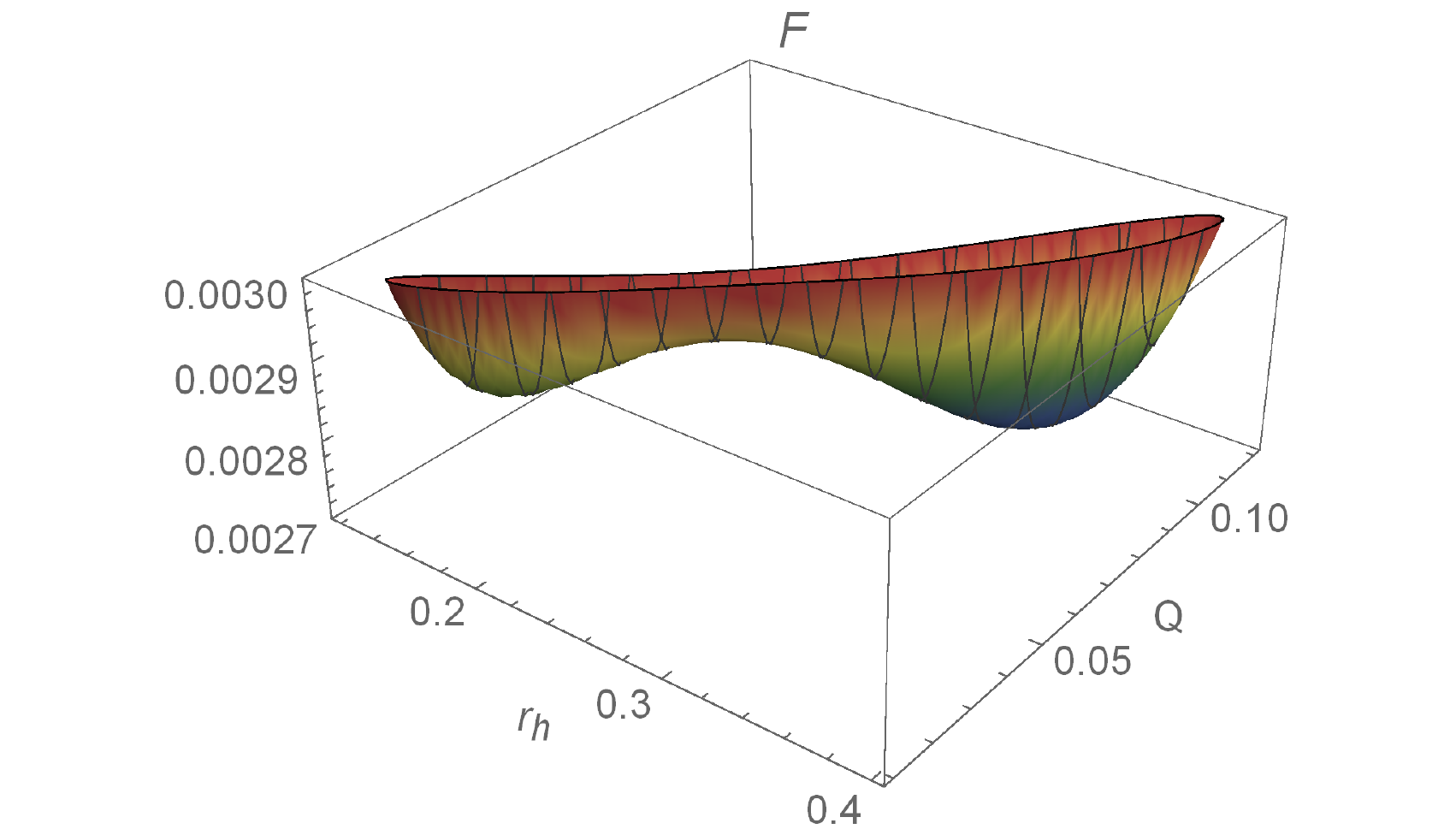}\\
  \caption{Two dimensional free energy landscape at the ensemble temperature that is lower (upper plot) and higher (bottom plot) than the critical temperature of phase transition.}
  \label{Free_Energy_Landscape_lowandhigh}
\end{figure}

One can also consider the case that the ensemble temperature is not equal to the critical temperature of phase transition. In Fig.\ref{Free_Energy_Landscape_lowandhigh}, the upper/bottom plot shows that the ensemble temperature is lower/higher than the critical temperature. In these two cases, the two basins on the landscape have different depths. The black hole state that is represented by the lowest point in the deeper basin is globally stable while the black hole state that is represented by the lowest point in the shallower basin is locally stable. From the plots, we can conclude that below the critical temperature the small black hole state is globally stable while above the critical temperature the larger black hole state is globally stable. The stabilities of the black holes are completely determined by the topography of the free energy landscape characterized by the basin depths and the barrier height between the basins. From the free energy landscape, one can easily read off the stabilities of the on-shell black holes.

\section{The topology and defects of the free energy landscape}
\label{topology}

In this section, we discuss the recent proposal of treating black holes as topological defects. Recently, based on the concept of the generalized free energy, the topological properties of the thermodynamic parameter space was investigated in Ref.\cite{Wei:2022dzw}. It is further argued that black hole can be treated as topological defect and the topological number can be used to characterize the stability of black hole. By introducing an ancillary parameter $\Theta$, the authors found some universal topological properties in the "extended" parameter $(r_h,\Theta)$ space. In particular, it is conjectured that the positive and negative winding numbers of the defects correspond to the local thermodynamical stable and unstable black holes. In this subsection, we will point out that there is an inconsistency in this proposal \cite{Liu:2022aqt,Fang:2022rsb}.

More recently, another method was introduced to study the intrinsic topological properties of black hole thermodynamics \cite{Bai:2022klw}. Using the spinodal $T-r_h$ curve, thermodynamic critical points of a black hole are endowed with the topological quantity of Brouwer degree. In addition, the topological transition between the different thermodynamic systems and the topological classification for them can be conveniently investigated. However, their analysis are not based on the free energy landscape. The relation between the topology of the free energy landscape and the stabilities of the on-shell black holes was not discussed.

\begin{figure}
  \centering
  \includegraphics[width=8cm]{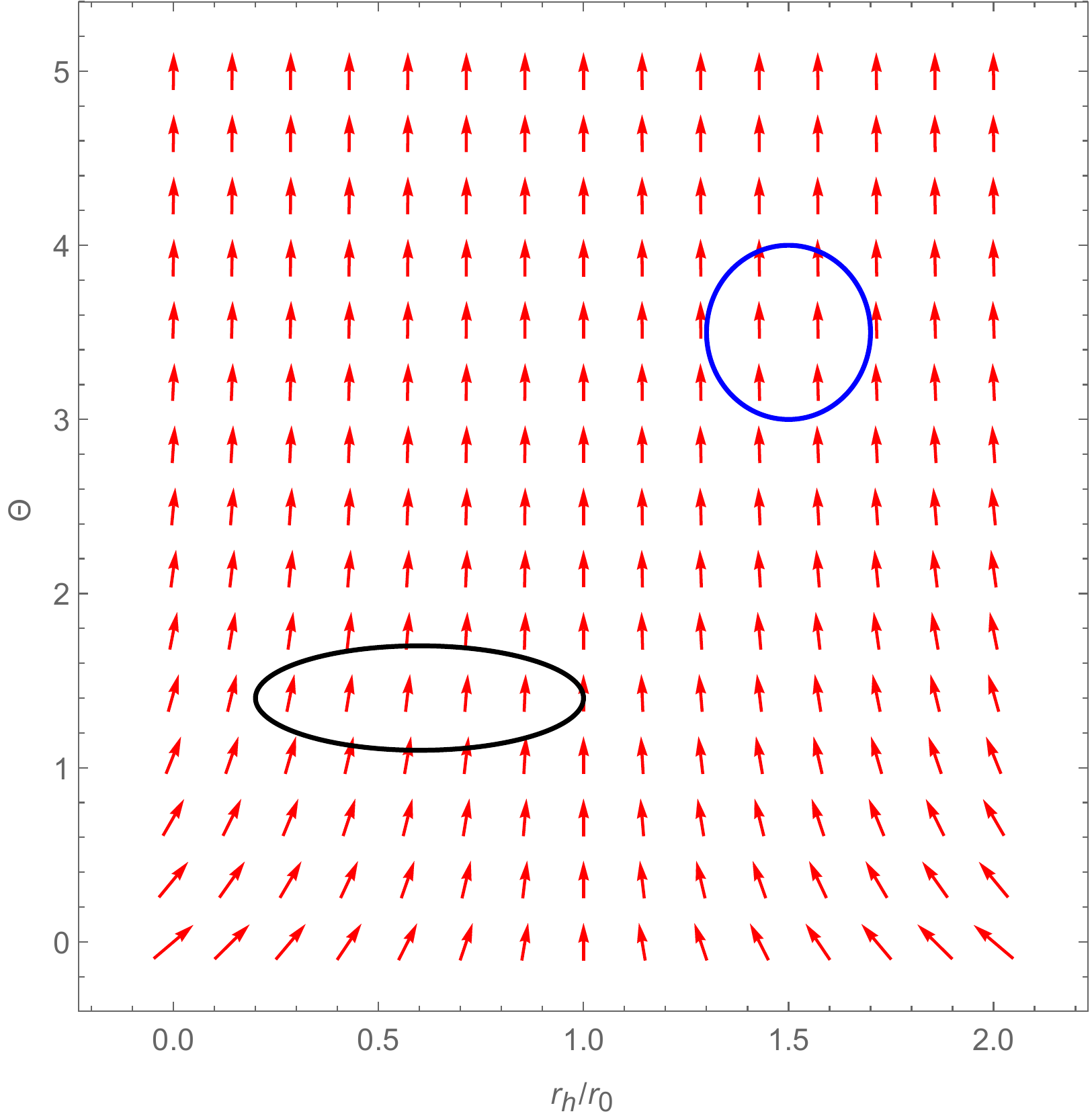}\\
  \caption{The plot of the vector field $\phi=(\frac{1}{2}-2\pi T r_h,e^{\Theta})$ for the Schwarzschild black hole. Comparing to the Fig.1 in Ref.\cite{Wei:2022dzw}, there is no topological defect in this plot. }
  \label{VectorPlot}
\end{figure}

Firstly, let us discuss whether the artificially introduced vector field $\phi$ in the $r_h-\Theta$ space can be used to describe the topology of the landscape intrinsically \cite{Wei:2022dzw}. Note that $\Theta$ is an ancillary parameter. The intrinsic topology should not depend on the explicit form of the introduced vector field $\phi$. It is obvious that, if we set the $\Theta$ component of the vector field $\phi$ as an arbitrary positive definite function, the vector field $\phi$ has no zero point on the whole parameter space. Fig.\ref{VectorPlot} shows the unit vector field $n^a=\phi^a/||\phi||$ for the Schwarzschild black holes when $\phi^{\Theta}$ is set to be $e^{\Theta}$. There is no topological defects on the whole thermodynamic parameter space. Then, the winding number for arbitrary loop is zero. Therefore, the topological defect and the calculation of the winding number by using the ancillary parameter $\Theta$ is not intrinsic. This conclusion is obviously valid for other types of black holes. The reason is that the generalized free energy $F$ in the canonical ensemble is intrinsically one dimensional and it only depends on the order parameter $r_h$. The introduced parameter $\Theta$ and the vector field $\phi$ have no significant physical meaning. It appears that the intrinsic topological properties of the parameter space cannot be revealed by the artificially introduced vector field $\phi$.

\begin{figure}
  \centering
  \includegraphics[width=8cm]{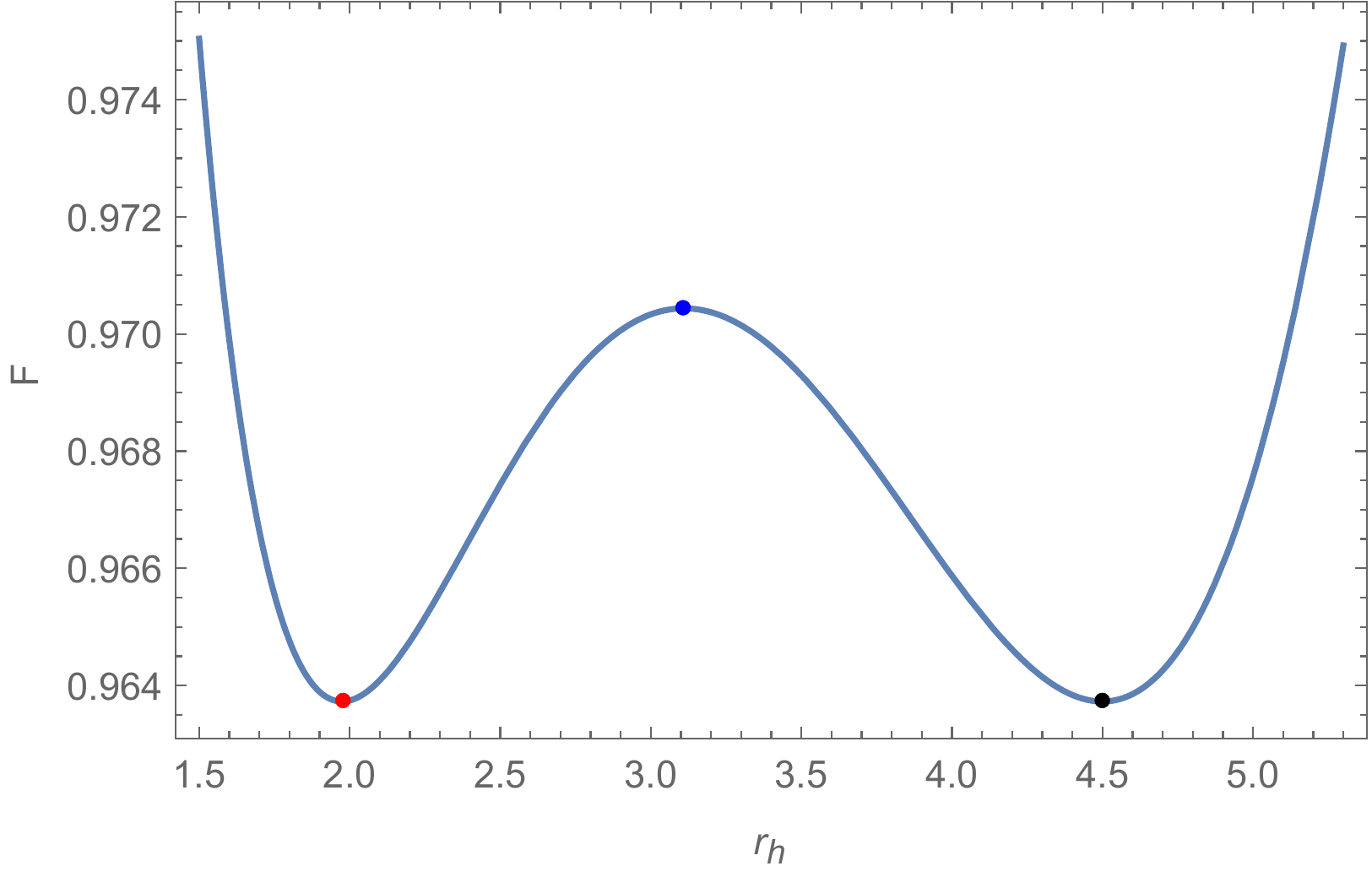}\;\;\;\;
  \includegraphics[width=8cm]{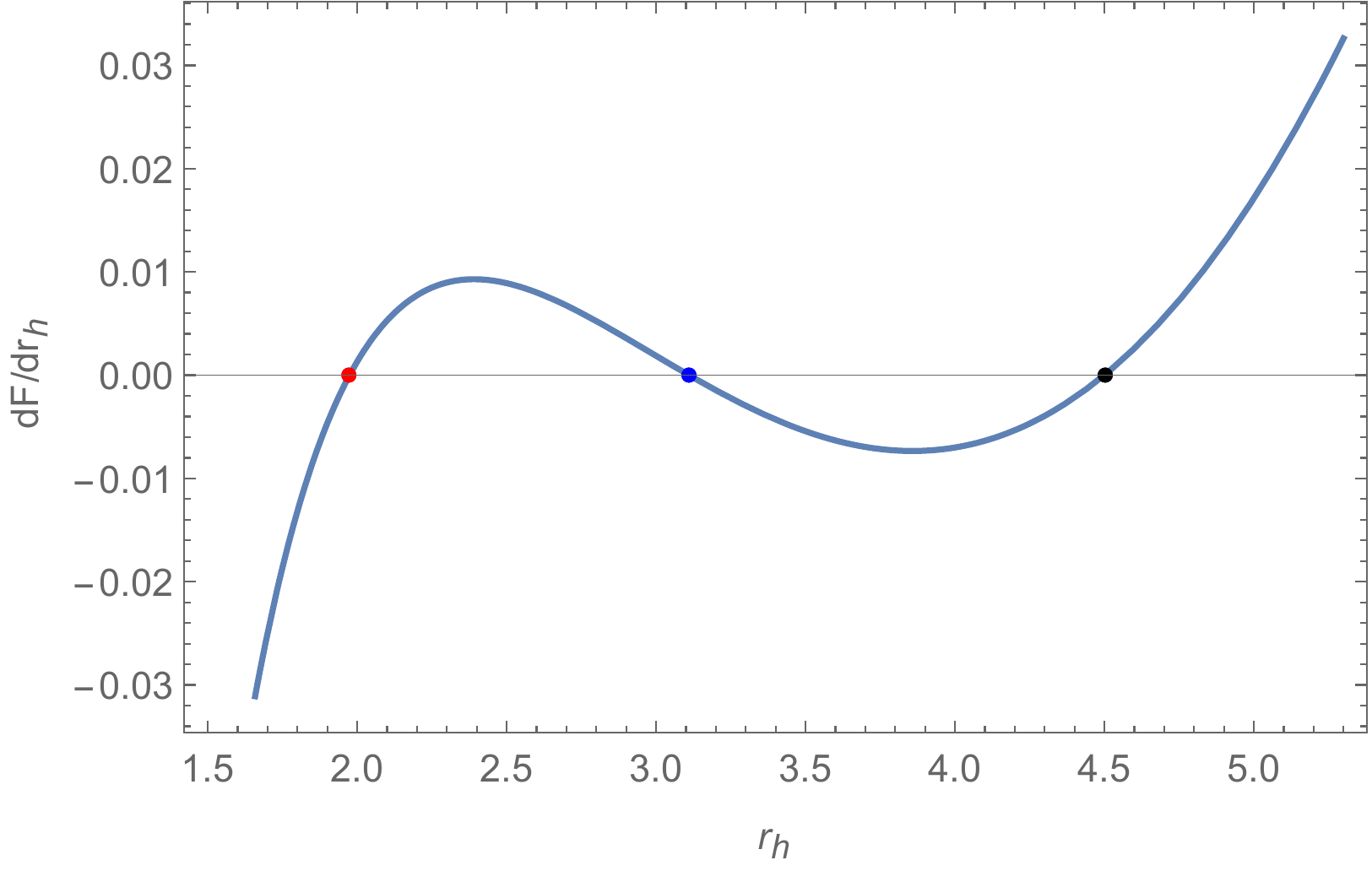}\\
  \caption{Upper panel: Generalized free energy for the RNAdS black holes at the phase transition critical point. Lower panel: The gradient of the generalized free energy. The red, blue and black points represent the small, intermediate and large black holes. They are also the singularity or the topological defects of the gradient field.}
  \label{Free_Energy_Landscape_RNAdS}
\end{figure}

Then, let us consider whether there is an intrinsic definition of topological defects based on the generalized free energy function. In canonical ensemble, the generalized free energy $F$ is the one variable function of the order parameter $r_h$. One can naturally introduce the gradient field $\phi=d F/d r_h$ to describe the intrinsic topology of the free energy function. As an illustration, we consider the case of RNAdS black holes. In Fig.\ref{Free_Energy_Landscape_RNAdS}, the free energy landscape and the gradient field of the RNAdS black hole are plotted. We can define an unit vector field as $n=\frac{d F/d r_h}{\left|d F/d r_h\right|}$. Notice that this unit vector has singularities (where $d F/d r_h=0$) on the landscape. Obviously, these singularities can be treated as the topological defects according to Duan's $\phi$-mapping theory \cite{Duan}. In addition, these singularities are just the on-shell black hole solutions. Therefore, black hole solutions can be treated as the topological defects of the gradient field of the free energy function, because the on-shell black hole solutions correspond to the zero points of the gradient field on the free energy landscape. Comparing the claim that black hole solutions are the topological defects for the artificially introduced field $\phi$, our definition of topological defects are more natural and intrinsic.

One can associate a topological charge with the singularity as following. Explicitly, one can define the index or Brouwer degree for the gradient $d F/d r$ as  \cite{Brouwer_Degree,Bai:2022klw}
\begin{eqnarray}
\textrm{ind}_{d F/d r_h}(r_h)=\frac{1}{2}\left(\textrm{Sign}[n|_{r_h+\epsilon}]-\textrm{Sign}[n|_{r_h-\epsilon}] \right)\;.
\end{eqnarray}
Note that the unit vector $n$ reflects the direction of the gradient of the landscape and the gradient changes the direction at the singularities. These singularities are just the basin or the barrier top on the landscape. For the basins, the index is $+1$. For the barrier top, the index is $-1$. The results are trivial for the one dimensional landscape. It seems that the index is related to the stabilities of the black holes because the basins are locally stable while the barrier top is unstable.

Let us consider a little bit more nontrivial example, i.e. the free energy landscape for the five dimensional Gauss-Bonnet AdS black holes in the grand canonical ensemble \cite{Cai:2013qga,Zou:2014mha}. The topological property of the free energy landscape can be naturally described by the gradient field of the surface, which is defined as $\phi=\left(\frac{\partial F}{\partial r_h},\frac{\partial F}{\partial Q}\right)$. It is easy to see that the on-shell black holes are all the critical points of the gradient field and also the singularities of the unit vector $n^a=\frac{\phi^a}{||\phi||}$. Once again, we observe that the black hole solutions can be treated as the topological defects of the gradient field of the landscape. This can be further explained by using Duan's $\phi$-mapping \cite{Duan} as follows.

With the two-dimensional vector field $\phi=(\phi^1,\phi^2)$, the topological current theory tells us that each zero point $z_i$ of $\phi$ can be endowed a winding number $W_i$. By introducing the topological current $j^{\mu}$ as 
\begin{eqnarray}
 j^{\mu}=\frac{1}{2\pi} \epsilon^{\mu\nu\rho}\partial_{\nu}n^a\partial_{\rho}n^b=\delta^2(\phi)J^{\mu}(\frac{\phi}{x})\;,  
\end{eqnarray}
where $x^{\mu}=(t, r_h, Q)$, it can be shown that 
\begin{eqnarray}
W&=&\int_{\Sigma}j^0 d^2 x=\sum_{i=1}^N W_i,
\end{eqnarray}
where $N$ is the number of the zero points of vector field $\phi$. 
The Jacobian vector $J^{\mu}(\frac{\phi}{x})$ is defined as 
\begin{eqnarray}
\epsilon^{ab}J^{\mu}(\frac{\phi}{x})=\epsilon^{\mu\nu\rho}\partial_{\nu}{\phi}^a\partial_{\rho}{\phi}^b. 
\end{eqnarray}
The above equations explicitly show that the zero points the gradient field $\phi=(\frac{\partial F}{\partial r_h}, \frac{\partial F}{\partial Q})$ of the generalized free energy landscape, which correspond to the on-shell Gauss-Bonnet AdS black holes, are just the defects of the topological current.

In the current case, the small and the large black holes are the sink points on the landscape, and the intermediate black hole is the saddle point. From the conclusion of the topology, one knows that the index of the saddle point is $-1$, while the index of the sink point is $+1$. Therefore, from the topology of the two dimensional free energy landscape, one can reach the conclusion that the index of the thermodynamic stable black holes is $+1$ while the index for the unstable black holes is $-1$. It seems that the index is related to the stabilities of the black holes.

Although the two cases we have considered support the conjecture that the topological numbers are related to the stabilities of black holes apparently, one can easily give a counterexample to this argument. Note that
\begin{eqnarray}
 \delta^2(\phi)=\sum_{i=1}^N \frac{1}{|J^0(\frac{\phi}{x})|_{z_i}}\delta(x^1-z^1_i)\delta(x^2-z^2_i),  
\end{eqnarray}
where $J^0(\frac{\phi}{x})$ can be explicitly expressed as 
\begin{eqnarray}
J^0(\frac{\phi}{x})=\frac{\partial^2 F}{\partial {r_h}^2}\frac{\partial^2 F}{\partial Q^2}-(\frac{\partial^2 F}{\partial r_h \partial Q})^2\;.
\end{eqnarray}
Then the winding number can be derived as
\begin{eqnarray}
W_i=\textrm{Sign}[J^0(\frac{\phi}{x})]|_{z_i}.
\end{eqnarray}
On the one hand, $J^0(\frac{\phi}{x})$ is the Jacobian between $(\phi^1, \phi^2)$ and $(r_h, Q)$. On the other hand, $J^0(\frac{\phi}{x})$ is the determinant of the Hessian matrix of the two-dimensional free energy $F(r_h, Q)$, where the local extremum points yield $J^0(\frac{\phi}{x})>0$ and the saddle points yield $J^0(\frac{\phi}{x})<0$. Namely, the stable minimums and the unstable maximums possess the same winding number $+1$, which is disobedient from the view that the positive or negative winding number corresponds to the stable and unstable black holes.

\begin{figure}
  \centering
  \includegraphics[width=8cm]{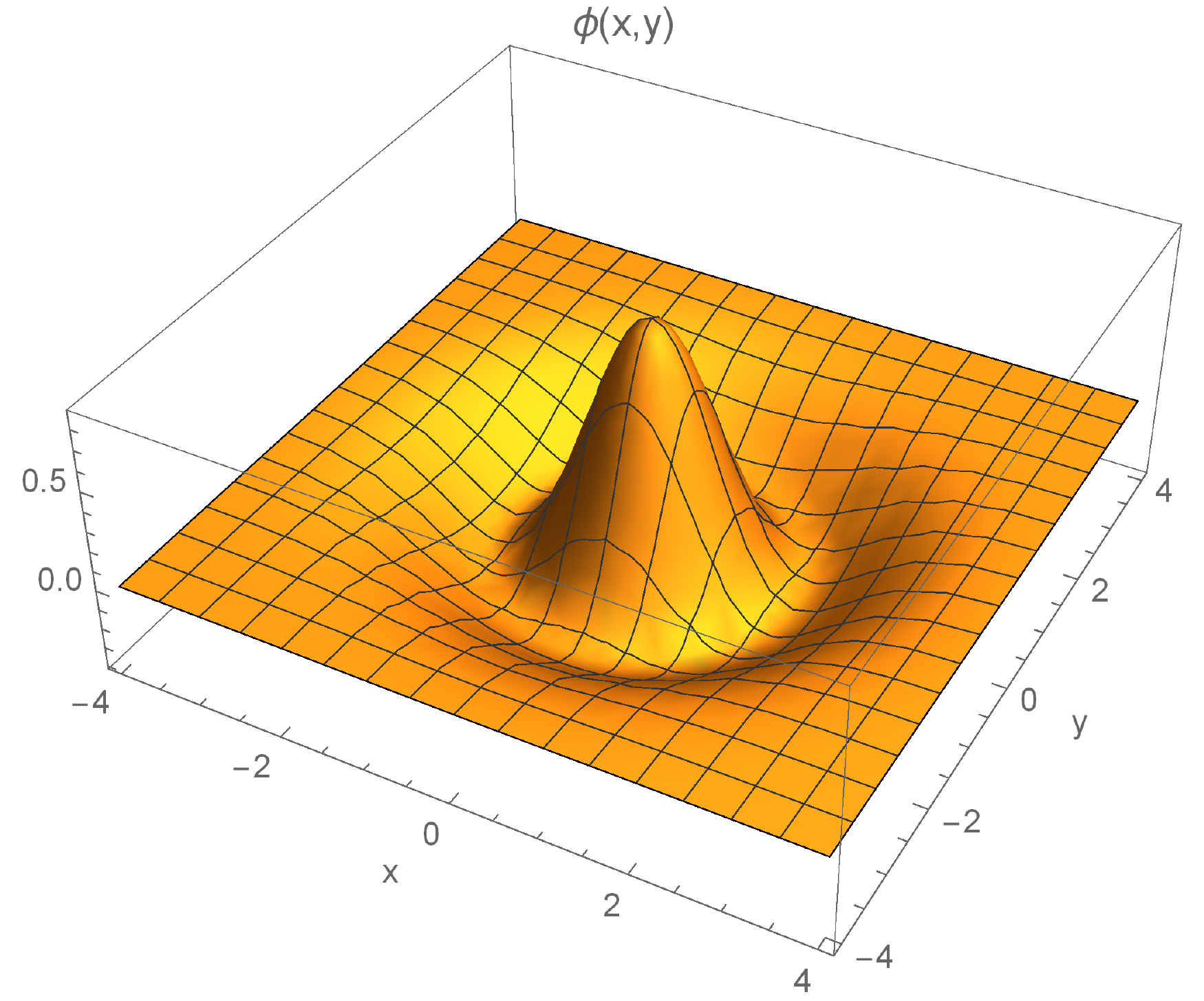}\\
  \caption{An illustration of the Mexican-hat potential $\phi(x,y)=-\frac{2 e^{-\frac{1}{2} \left(x^2+y^2\right)} \left(x^2+y^2-1\right)}{\sqrt{3} \sqrt[4]{\pi }}$. }
  \label{Mexican_hat}
\end{figure}

The above explanation shows that, in the two dimensional landscape, there may exist source, sink/basin, and saddle points (see Fig.13(b) in \cite{JW} for an illustration). Only the sink/basin points are thermodynamically stable. The source (the top of the mountain) and saddle points are unstable. On the other hand, from the knowledge of topology, the winding numbers of the source and the sink/basin are all $+1$, and for the saddle point, the winding number is $-1$. To be more clear, it is instructive to consider the Mexican-hat potential as illustrated in Fig.\ref{Mexican_hat}. The state that represented by the top of the potential is unstable while the winding number is $+1$. This simple example shows that the topological number-wingding number can not be used to characterize the stability of the states on the free energy landscape in general. For a more concrete example, one can refer to the non-equilibrium free energy landscape that presented in Fig.13(b) in Ref.\cite{JW}. Therefore, the topology is not relevant to the stabilities of thermodynamic systems in this setup. We should emphasize that the thermodynamics stabilities of the black holes are completely determined by the topography not the topology of the free energy landscape.

\section{Dominant kinetic path of the stochastic transition on the landscape}
\label{kinetics}

In this section, we discuss the stochastic motion of the local stable black hole state on the landscape by taking the thermal and charge fluctuations into account. We will quantify the dominant kinetic path of the phase transition from one stable black hole state to another.

\subsection{Stochastic dynamics of black hole state on the landscape}

We have shown that the local stable black holes are represented by the sink points on the free energy landscape. Without the fluctuations, an unstable state deviated from the local stable point will return to the local stable state under the driving force provided by the landscape. In this case, no phase transition will happen because the thermodynamic driving force is a deterministic force that pulls the unstable state to the valley bottom on the landscape.

Considering the black hole as a thermal entity, there are fluctuations from the external fluctuations of the bath and the intrinsic statistical fluctuations. These fluctuations are the results of the interactions between the degrees of freedom of the black hole and the degrees of freedom of the bath. These fluctuations will in turn result in the stochastic motion of the black hole state on the landscape. For simplicity, we will use $r$ to denote $r_h$. We propose that the stochastic motion of the black hole state is described by the Langevin equation that determines the stochastic evolution of the black hole order parameters $(r,Q)$ as follows \cite{Li:2021vdp} 
\begin{eqnarray}
\frac{d \vec{X}}{dt}=\vec{f}+\vec{\eta}\;,
\end{eqnarray}
where $\vec{X}=(r,Q)$ is the vector composed by the order parameters. In the above equations, $\vec{f}=-\tensor{D}\cdot \nabla (\beta F)$ is the driving force which is given by the gradient of the landscape $F(r,Q)$, and $\vec{\eta}$ is Gaussian noise term where its correlation is $\langle \vec{\eta}(\vec{X},t) \vec{\eta}(\vec{X},t') \rangle=2 \tensor{D} \delta(t,t')$ with $\tensor{D}$ being the diffusion coefficient matrix. In the present work, we take the diffusion coefficient matrix $\tensor{D}$ as the diagonal constant matrix $\textrm{Diag}\{D_r,D_Q\}$. In general, $D_r\neq D_Q$, which represents the inhomogeneity of the diffusion along the different directions of the order parameters. Then the driving force $\vec{f}=\left(-\beta D_{r}\partial_r F ,-\beta D_{Q} \partial_Q F \right)$ with $\beta$ being the inverse temperature.

It is well known that the stochastic dynamics can be formulated in terms of the path integral. The probability of starting from an initial configuration $\vec{x}_0$ at $t=0$ and ending at a final configuration $\vec{x}$ at time $t$ is given by the Onsager–Machlup functional \cite{Onsager,Hanggi}
\begin{eqnarray}
P(\vec{x},t;\vec{x}_0,0)=\int \left[\mathcal{D}\vec{X}\right] \textrm{exp}\left\{-\int dt \left[\frac{1}{2}\nabla\cdot \vec{f} + \left(\frac{d \vec{X}}{dt}-\vec{f}\right) \cdot \frac{1}{4\tensor{D}} \cdot \left(\frac{d \vec{X}}{dt}-\vec{f}\right) \right]\right\}\;,
\end{eqnarray}
where $1/\tensor{D}$ means the inverse matrix of the diffusion coefficient matrix. The integral over $\mathcal{D}\vec{X}$ denotes the sum over all possible paths from the state $\vec{x}_0$ at time $t = 0$ to the state $\vec{x}$ at time $t$, and the exponential factor gives the weight for each path. Therefore, the path integral is the sum of the weights of all possible paths and can be approximated with a set of dominant paths. The dominant kinetic path gives the path of the stochastic state transition process on the landscape with the highest optimal weight.

To proceed, one can introduce the action for the path integral as \cite{WangJCP2010}
\begin{eqnarray}
S=\int dt \mathcal{L}\left(\vec{X}(t)\right)=\int dt \left[\left(\frac{d \vec{X}}{dt}-\vec{f}\right) \cdot \frac{1}{4\tensor{D}} \cdot \left(\frac{d \vec{X}}{dt}-\vec{f}\right)+\frac{1}{2}\nabla\cdot \vec{f} \right]\;.
\end{eqnarray}
The dominant kinetic path with the optimal weight can be obtained by minimizing the action or Lagrangian. The Lagrangian can be explicitly written as  
\begin{eqnarray}
\mathcal{L}=\frac{1}{4D_{r}}\left( \frac{dr}{dt} -f_r \right)^2+\frac{1}{4D_{Q}}\left( \frac{dQ}{dt} -f_Q \right)^2 +\frac{1}{2}\left( \partial_r f_r+\partial_Q f_Q \right)\;.
\end{eqnarray}
The corresponding Euler-Lagrangian equations for the dominant paths are given by 
\begin{eqnarray}\label{EL_eq}
\frac{1}{2\tensor{D}} \frac{d^2 \vec{X}}{dt^2}=\nabla V(\vec{X})\;,\;\;\;
V(\vec{X})=\vec{f}\cdot \frac{1}{4\tensor{D}}\cdot \vec{f}+\frac{1}{2}\nabla\cdot \vec{f}\;,
\end{eqnarray}
or in the component form as 
\begin{eqnarray}
\frac{1}{2D_{r}}\frac{d^2 r}{dt^2}&=& 
\frac{1}{2D_{r}} f_r \partial_r f_r +\frac{1}{2D_{Q}} f_Q \partial_r f_Q +\frac{1}{2} 
\left(\partial_r^2 f_r+\partial_r\partial_Q f_Q\right)\;,\\
\frac{1}{2D_{Q}}\frac{d^2 Q}{dt^2}&=& 
\frac{1}{2D_{r}} f_r \partial_Q f_r +\frac{1}{2D_{Q}} f_Q \partial_Q f_Q +\frac{1}{2} 
\left(\partial_r\partial_Q f_r+\partial_Q^2 f_Q\right)\;.
\end{eqnarray}
It is easy to check that the conserved equation is 
\begin{eqnarray}\label{conserved_eq}
E=\frac{1}{4\tensor{D}}\left(\frac{d\vec{X}}{dt}\right)^2-V(\vec{X})\;,
\end{eqnarray}
where $E$ is a conserved quantity. This equation can be viewed as a particle with the effective mass $\frac{1}{2\tensor{D}}$ moving along the two dimensional effective potential $-V(\vec{X})$.

In principle, one can solve the Euler-Lagrangian equations with the fixed boundary conditions to obtain the dominant kinetic path for the stochastic motion of the black hole state on the landscape. However, this is a numerically challenging problem when dealing with the two end boundary conditions especially for high dimensions. In the following, we invoke the Hamiltonian-Jacobian approach \cite{Landau} to obtain the dominant kinetic paths.

\subsection{Dominant kinetic path from Hamiltonian-Jacobian method}

We focus on the most probable path with the maximal contribution to the Onsager-Machlup functional, which means that the exponential weight $e^{-S}$ is maximum and hence the action $S$ is minimum. In addition, we have observed that the effective dynamics is conserved. This problem can be solved by using Maupertuis’ principle in classical mechanics \cite{Landau,Wangprl2006,Faccioli2006,WangJCP2010}, which is about the physical path that connects given initial and final positions with the initial energy $E$ fixed. Different from Hamilton's principle, where the path as the function of time is determined, Maupertuis's principle determines only the shape of the path. This is to say that we will switch from the time-dependent Newtonian description to the energy-dependent Hamilton-Jacobi description \cite{Wangprl2006,Faccioli2006,WangJCP2010}.  

Specifically, we want to identify the dominant kinetic path that minimizes the action $S$. As stated, we can use the Maupertuis's principle to solve this problem \cite{Wangprl2006,Faccioli2006,WangJCP2010}. Maupertuis's principle states that the dominant kinetic path connecting given initial and final positions with fixed energy $E$ is obtained by minimizing the abbreviated action functional \cite{Landau} 
\begin{eqnarray}
S_0=\int_{\vec{x}_i}^{\vec{x}_f} \vec{P}\cdot d\vec{X}\;,
\end{eqnarray}
where $\vec{P}$ is the generalized canonical momentum 
\begin{eqnarray}
\vec{P}=\frac{\partial \mathcal{L}}{\partial \dot{\vec{X}}}=
\frac{1}{2\tensor{D}} \cdot \left(\frac{d \vec{X}}{dt}-\vec{f}\right)\;.
\end{eqnarray}
Note that it does not always work for arbitrary energy value $E$. But what we are considering is the path that minimizes the action $S_0$ not the time that the particle takes from the initial position to the final position. The final time that the particle arrives at the final position is flexible which means that there will be many valid physical paths for a wide range of energy values.

By using the conserved equation (\ref{conserved_eq}), the abbreviated action functional can be rewritten as 
\begin{eqnarray}
S_0=\int_{\vec{x}_i}^{\vec{x}_f} \sqrt{(E+V(\vec{X}(l)))/D} dl\;,
\end{eqnarray}
where $dl$ is an infinitesimal displacement along the path trajectory. Note that because the driving force $\vec{f}$ is a gradient force, its integral is path independent. Its contribution to the action has been omitted in the expression of the action functional. In fact, $E$ is a free parameter that determines the total kinetic time of the transition process. In the present work, we adopted the simple choice $E=\textrm{Maximum}\{-V(\vec{x}_i),-V(\vec{x}_f)\}$, which corresponds to the longest kinetic time.

In practice, we should firstly discretize the abbreviated action by dividing the path into $N$ steps \cite{Faccioli2006,WangJCP2010}. Then the discretized action can be written as 
\begin{eqnarray}
S_0=\sum_{n=0}^{N-1} \sqrt{(E+V(n)/D} \Delta l_{n,n+1}+\lambda P\;,   
\end{eqnarray}
where $P$ is a penalty function 
\begin{eqnarray}
P=\sum_{n=1}^{N-1}\left( \Delta l_{n,n+1}-\langle \Delta l \rangle   \right)^2\;, 
\end{eqnarray}
which is introduced to keep all the length elements close to their average and irrelevant in the continuum limit. The minimization of the discretized abbreviated action can be performed by applying both simulated annealing and conjugate gradient algorithm \cite{Faccioli2006,WangJCP2010}. Finally, the dominant kinetic path can be visualized by projecting the numerical results to the order parameter $(r_h,Q)$ space. The weight for the dominant kinetic path can be obtained by substituting the path into the abbreviated action.

\begin{figure}
  \centering
  \includegraphics[width=8cm]{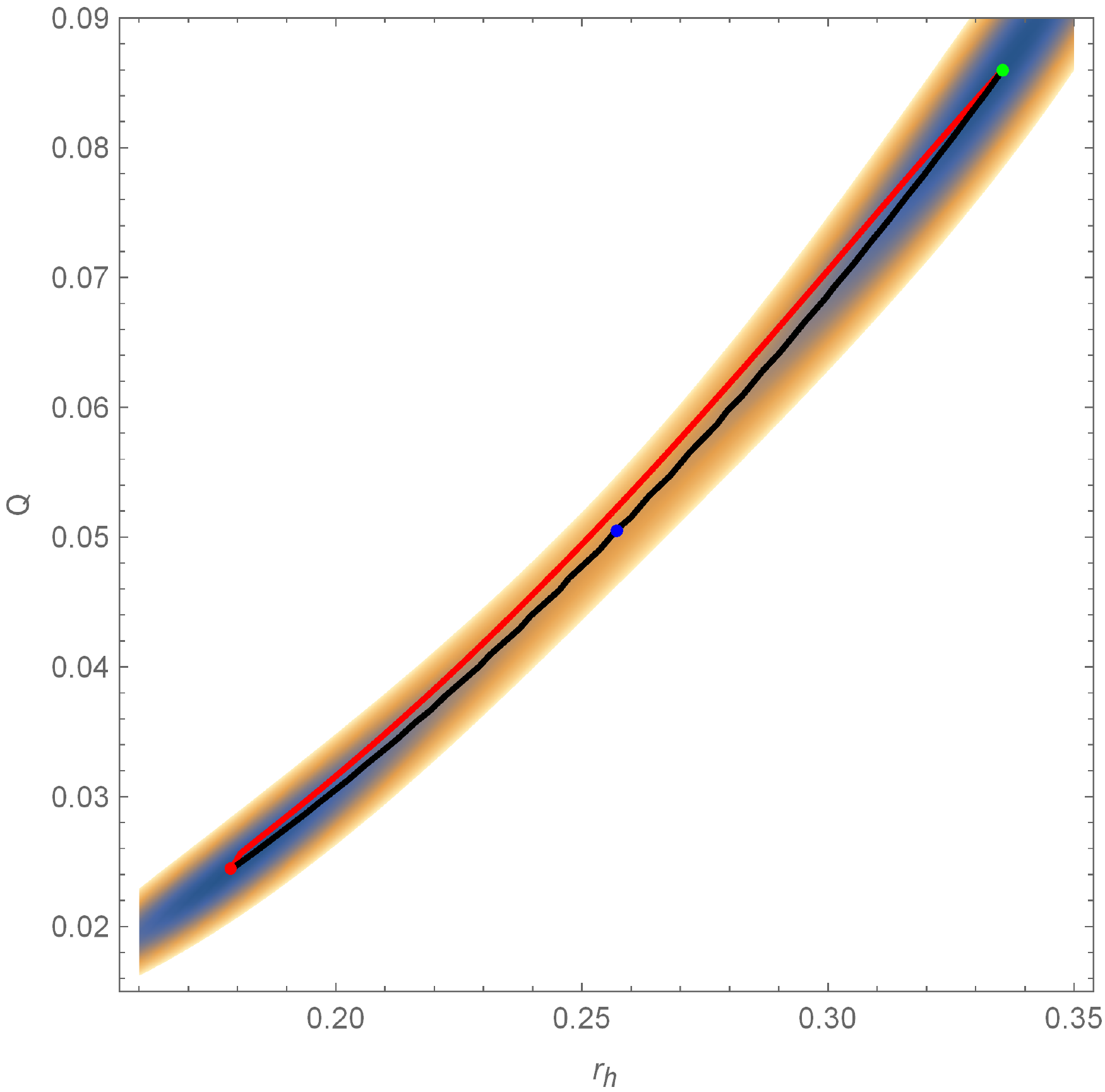}
  \includegraphics[width=2.2cm]{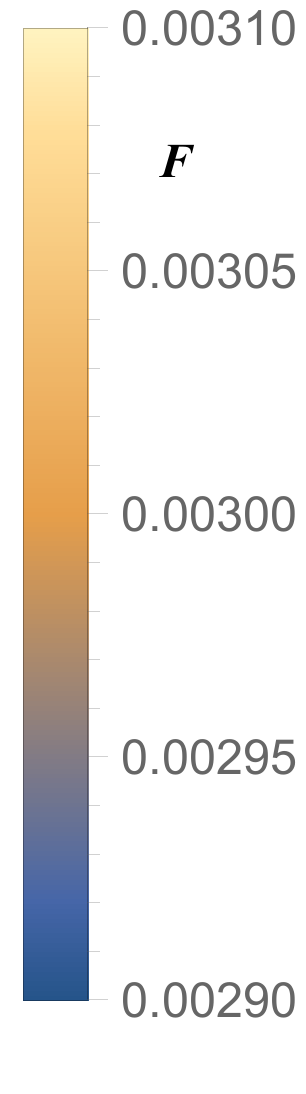}\\
  \caption{Dominant kinetic paths of the phase transition process on the free energy landscape for the diffusion coefficients at zero limit (black line) and at finite value (red line). In this plot, the black line and the yellow line are obtained by setting $D_r=D_Q=10^{-8}$ and $D_r=D_Q=0.005$, respectively. }
  \label{paths_zeroDlimit_vs_finite}
\end{figure}

\begin{figure}
  \centering
  \includegraphics[width=8cm]{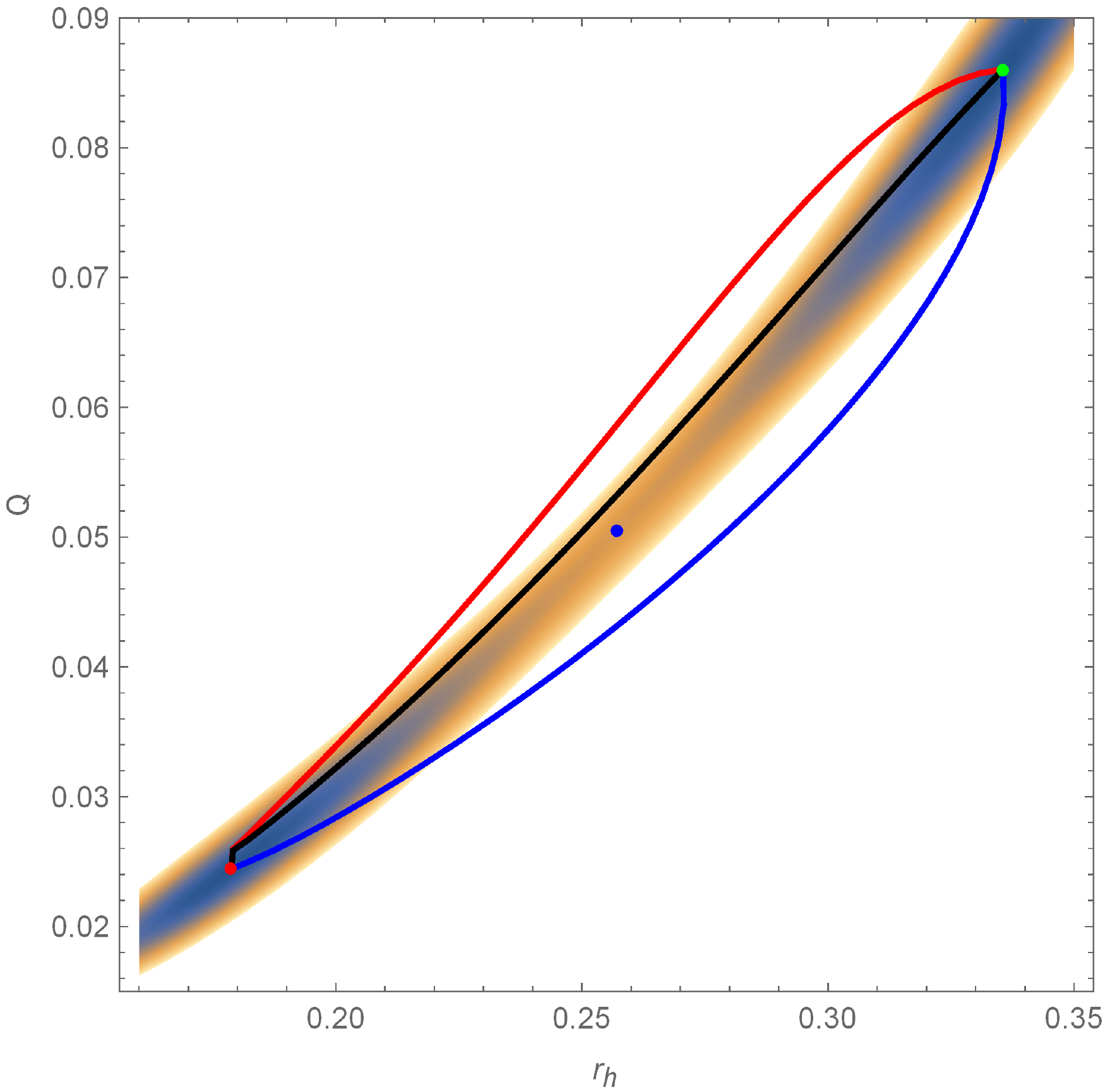}
  \includegraphics[width=2.2cm]{path_zeroDlimit_vs_finite_1.pdf}\\
  \caption{Dominant kinetic paths for the phase transition on the free energy landscape. The red, black and blue lines correspond to the diffusion coefficients $(D_r,D_Q)=(0.004,0.008)$, $(D_r,D_Q)=(0.008,0.008)$, and $(D_r,D_Q)=(0.016,0.0016)$, respectively. }
  \label{paths}
\end{figure}

The numerical results are presented in Fig.\ref{paths_zeroDlimit_vs_finite} and Fig.\ref{paths}, where the dominant kinetic paths for different diffusion coefficients are plotted. In Fig.\ref{paths_zeroDlimit_vs_finite}, the dominant kinetic paths for the diffusion coefficients $D$ at the zero limit (black line) and at finite value (red line) are plotted in the order parameter $(r_h,Q)$ space. It is shown that when the diffusion coefficient is very small, the dominant kinetic path for the transition between the small and the large Gauss-Bonnet black holes passes through the intermediate black hole state. The diffusion coefficient $D$ in the zero limit corresponds to the case when the fluctuations are very small. In this case, the term $\nabla \cdot \vec{f}$ in the effective potential $V(\vec{X})$ in Eq.(\ref{EL_eq}) can be ignored, then the dominant path does pass through the saddle point. When the diffusion coefficients are finite, the fluctuations are also finite. Then the contribution from the term $\nabla \cdot \vec{f}$ can be significant and the resulting dominant kinetic path does not necessarily pass through the saddle point. The red line in Fig.\ref{paths_zeroDlimit_vs_finite} shows that when $D_r=D_Q=0.005$, the dominant kinetic path of the transition process does not pass through the intermediate Gauss-Bonnet black hole state. This is very different from the kinetic path for the one dimensional landscape \cite{Liu:2021lmr}, where the kinetic path between the small black hole state and the large black hole state must pass through the intermediate black hole state.

In Fig.\ref{paths}, the black line is the optimal path that dominants the action functional when $D_r=D_Q$. It is clear that this dominant kinetic path originates from the large Gauss-Bonnet black hole state, and ends on the small Gauss-Bonnet black hole state, but does not pass through the intermediate black hole state. The red and blue lines show that dominant kinetic paths that correspond to the diffusion coefficients $D_r<D_Q$ and $D_r>D_Q$. We can observe the same conclusion that the dominant kinetic path does not necessarily pass through the intermediate black hole state. In addition, when there is a contrast between the diffusion coefficients characterizing the degrees of fluctuations in different directions in order parameter space, the dominant paths bias towards the higher diffusion direction the first and move along the lower diffusion direction the last. This shows that the inhomogeneity in diffusions can lead to the switching from the coupled cooperative process of black hole phase transition between $r_h$ and $Q$ along the diagonal line in the order parameter space to the decoupled sequential process with first moving in $r_h$ direction and then $Q$ direction, or first moving in $Q$ direction and then $r_h$ direction. This gives rise to the new kinetic mechanisms ranging from cooperative to sequential process for the black hole phase transitions.

\section{Conclusion}
\label{conclusion}

In summary, we have studied an example that the free energy landscape is an intrinsic two dimensional surface. In this example, the black hole radius and charge are selected to be the order parameters, and the generalized free energy of the five dimensional charged Gauss-Bonnet black holes is defined properly in the grand canonical ensemble.

The landscape as the quantitative representation of the generalized free energy function indicates the stability of the on-shell black holes. We have emphasized that the stability of black hole is completely determined by the topography of the free energy landscape. On the landscape, we showed that the on-shell Gauss-Bonnet black holes are the extremum points and other points represents the off-shell black holes. In our case, there are three branches of charged Gauss-Bonnet black hole solutions, which are classified by their black hole radii. The global/local stable black holes are at the lowest points in the basins while the unstable black hole is at the saddle point. We also comment on the recently proposal of viewing black hole as topological defect. We argued that the black hole can be treated as topological defect of the gradient field of the landscape, but the stability is not necessarily related to the topology of the gradient field of the landscape.

In addition, we studied the stochastic dynamics of the black hole phase transition and quantified the dominant kinetic path for the state transition on the two dimensional free energy landscape. We showed that the dominant kinetic path passes through the intermediate black hole state when the fluctuation is very small while at the finite fluctuations it does not necessarily pass through the intermediate black hole state. The inhomogeneity in diffusions can  switch the coupled cooperative process of black hole phase transition to the decoupled sequential process, leading to different kinetic mechanisms.

\end{document}